# A USE-CASE DRIVEN APPROACH IN REQUIREMENTS ENGINEERING : THE MAMMOGRID PROJECT

Mohammed Odeh, Tamas Hauer, Richard McClatchey & Tony Solomonides
Centre for Complex Cooperative Systems, CEMS Faculty, University of the West of England,
Coldharbour Lane, Frenchay, Bristol BS16 1QY, United Kingdom
Telephone: +44 117 344 3700, FAX: +44 117 344 3155, Email: Mohammed.Odeh@uwe.ac.uk

## Abstract

We report on the application of the use-case modeling technique to identify and specify the user requirements of the MammoGrid project in an incremental and controlled iterative approach. Modeling has been carried out in close collaboration with clinicians and radiologists with no prior experience of use cases. The study reveals the advantages and limitations of applying this technique to requirements specification in the domains of breast cancer screening and mammography research, with implications for medical imaging more generally. In addition, this research has shown a return on investment in use-case modeling in shorter gaps between phases of the requirements engineering process. The qualitative result of this analysis leads us to propose that a use-case modeling approach may result in reducing the cycle of the requirements engineering process for medical imaging.

## 1. Introduction

In this paper, we report on the application of the use-case modeling technique to identify and specify the user requirements of the MammoGrid project in an incremental and controlled iterative approach. A brief overview of the MammoGrid [1] project with emphasis on the user requirements specifications is presented in section 2. Section 3 presents an introduction to requirements engineering processes and the utilization of the requirements workflow (of Rational's Unified Process [2]) in the elicitation and specification of the user requirements. The MammoGrid use-case model is presented in section 4 and a discussion on the impact of the use-case approach is presented in section 5 followed by a summary of the main outcomes in section 6.

## 2. The MammoGrid Project

The MammoGrid project aims to investigate the feasibility of developing a European database of mammograms, accessed using emerging Grids [3] software, so that a set of important healthcare applications using this database can be enabled and Grids can be harnessed to support co-working between healthcare professionals across the EU.

The main output of the MammoGrid project, a Grid-enabled software platform (called the MammoGrid Information Infrastructure) which federates multiple mammogram databases, will enable clinicians to develop new common, collaborative approaches to the analysis of mammograms. This will be achieved through the use of Grid-compliant services for managing massively distributed files of mammograms, for handling the distributed execution of mammograms analysis software, for the development of Grid-aware algorithms and for the sharing of resources between multiple collaborating medical centres. All this is delivered via a novel software and hardware information infrastructure that guarantees the integrity and security of the medical data.

The MammoGrid project is being driven by the requirements of its user community (represented by Udine (Italy) and Cambridge (UK) hospitals along with medical imaging expertise from Oxford) that will be elicited and specified in detail. Inside the first six months of the project a user requirements analysis has been undertaken by identifying and specifying the functional and non-functional requirements of the end-user radiologists and radiographers (radiologic technologists); by developing suitable system models such as use-case models, object models and interaction diagrams; by defining hardware and software requirements; and by the development of a logical view of the application architecture of the MammoGrid. The key deliverable of this study was the User Requirements Specification (URS) produced in Work package 2 (WP2) of the project (see figure 1).

The requirements have been specified by a group of software engineers (from the University of the West of England) working with domain experts (from Mirada Solutions [4] and the hospitals in Cambridge and Udine). In carrying out this specification, a requirements engineering process has been applied to elicit the requirements of the MammoGrid, to analyze these requirements, to develop requirements definitions and specifications, and to validate these requirements.

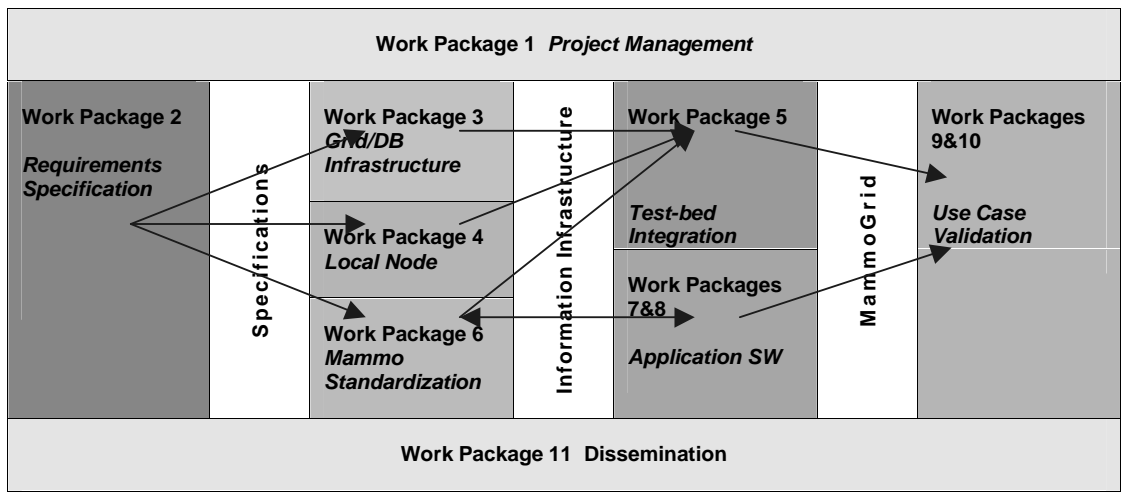

**Figure 1**: Workpackages of the MammoGrid Project

Rational's Unified Process model (RUP [2]) has been used in the requirements specification for MammoGrid and in particular key requirements engineering activities. The process has identified major use-case scenarios (in a Use Case Model) in the use of a distributed database of mammograms deployed across a pan-European Grid and that later can be used to prove the MammoGrid prototype.

To facilitate requirements specification, a number of meetings took place between software engineers and radiologists at Udine, Torino and Cambridge. Requirements were thus elicited for subsequent development into a requirements model. Problem domain objects were identified and described in addition to documenting relationships between such objects, resulting in a logical object model for MammoGrid. A number of other models including sequence and state transition diagrams, to model the dynamic behavior of MammoGrid, were also developed.

An analysis of system non-functional requirements (NFRs) was conducted including product-related, organization and process requirements, external constraints such as confidentiality and interface specifications such as Computer Aided Detection (CADe). In this analysis constraints on the process of mammogram study, such as usability, reliability, robustness and security were investigated and specified to assess the adherence to these requirements at implementation time. In addition, the impact of product-related NFRs on the selection and specification of the anticipated architecture of the MammoGrid was investigated.

The resulting MammoGrid User Requirements Specification (URS) details two essential objectives that must be supported and tested in the MammoGrid project:

- The support of clinical research studies through the access to and execution of algorithms on physically large, geographically distributed and potentially heterogeneous sets of (files of) mammographic images, just as if these images were locally resident.

- The controlled and assured access of educational and commercial companies to distributed mammograms for the purposes of testing novel medical imaging diagnostic technologies in scientifically acceptable clinical trials that fulfill the criteria of evidence-based medical research.

In relation to evidence-based radiological practice, the work of Bui and colleagues [3] came to our attention after this requirements specification was completed. We hope to address their analysis in a subsequent publication.

## 3. The Requirements Engineering Process

The requirements engineering process has been defined [6] as the set of activities that deal with problem domain understanding, requirements elicitation, requirements analysis, requirements definition and specification, requirements validation and requirements change management in an incremental and iterative manner. However, the application and enactment of certain requirements engineering processes may reasonably be expected to vary from project to project, and organization to organization [7].

The MammoGrid project has certain non-standard characteristics: a wide diversity of backgrounds among problem domain specialists (radiologists, radiographers, epidemiologists, medical imaging experts), the application domain itself, roughly speaking, the construction of the evidence base for radiological practice in mammography, and the geographically dispersed locations of the different parties involved in the project. In addition, the use of a modeling language such as UML [8] (and in particular the utilization of techniques from software development, such as use-cases) with corresponding modeling, validation, and requirements management tools, are normally linked to a software development life cycle model used throughout an organization or project. Here there was a need to establish these *ab initio*.

We used the Requirements Workflow of Rational's Unified Process [9] in guiding the requirements engineering activities in WP2. While use cases provided a common language for software engineers and domain experts to communicate in, and in particular for radiologists to translate their protocols into a somewhat more declarative form, other benefits accrued. Following this workflow has contributed to the clearer delineation of the scope of the project, by identifying its relationship with the external world through the use-case model described in section 4. Moreover, this view of the interaction between radiologist and MammoGrid led more or less directly to the specification of the user-interface of the software and to prioritization of the requirements. This has allowed the team to manage the scope of the project to advantage at a very early stage. Equally, the radiological and epidemiological scope of the project needed clear definition. Discussion of the use cases and the associated scenarios, data and communication requirements also led to appropriate definitions of the final work packages which will effectively 'prove' the project.

As with any other modeling technique, use case modeling restricts the scope of the modeler's thinking by ruling certain things out of consideration and by ruling only certain kinds of conceptual objects in. For one thing, because of its essentially declarative nature, certain natural elements (e.g. conditions) have to wait for the right place to be expressed. This is somewhat unnatural to the domain expert who wishes to state that (say) an extension of an action may only take place in certain circumstances, i.e. that this extension is not universally applicable. Actors' names also provided an interesting occasion for detailed discussion: a radiologist is a 'mammogram analyst' in MammoGrid because that is the radiologist's role in the system. These highly specialized physicians had to compartmentalize their work in order to make sense of the roles in the model.

## 4. The MammoGrid Use-Case Model

The User Requirements Specifications (URS) document identifies and specifies the functional and non-functional requirements of end-user radiologists and radiographers by developing system models such as use-case models, class models, and interaction diagrams; by defining software and hardware requirements; and by the development of the abstract 'logical' view into a 'deployment' model of the application architecture of the MammoGrid.

The choice of a use-case driven approach to requirements gathering seemed most practical as the gap between the developers' background knowledge of the domain and that of the users needed to be bridged in order to tackle the problem of architectural and interaction design. The requirements elicitation process was carried out in consultation with the user community at the two hospitals in Udine and Cambridge, with contributions from a further private hospital in Torino. In these discussions we defined nine core use cases with corresponding actors (one actor's role was eventually split between two use cases thus we defined eight actors), as shown in Figure 2. The refinement of this picture: extension of sub-use cases and the definition of the flows of each use case was then carried out as we identified all functional and non-functional requirements. The following is a brief summary of the core use cases.

*Use-Cases 1 and 2:* Maintain Mammogram X-rays and Maintain Patient Basic Details. These use-cases specify what the 'Database Operative' actor, normally a radiographer, undertakes in order to maintain the mammograms and corresponding patient data in the MammoGrid database.

*Use-Case 3:* Radiological Analysis. This use-case describes the tasks that the 'Mammogram Analyst' actor, normally a radiologist or perhaps an epidemiologist, may undertake to annotate and/or view mammograms and patient details, to execute radiological queries, including use of CADe, and to execute epidemiological queries.

*Use-Cases 4, 5 and 6:* Quality Control and Assurance. Three actors were identified whose role is related to quality control. The Quality Controller has primary responsibility for the maintenance and settings of all mammographic equipment; the Quality Assurance Officer supervises these controlled machines and reviews their maintenance. The Quality Control Research use-case describes the means by which to analyze the ways in which Standard Mammogram Form (SMF) can contribute to improving quality control in taking mammograms, including development of quantitative measurements of the quality of a mammogram. The results will support grid-enabled monitoring of the quality of mammograms submitted at each contributing centre to the MammoGrid database.

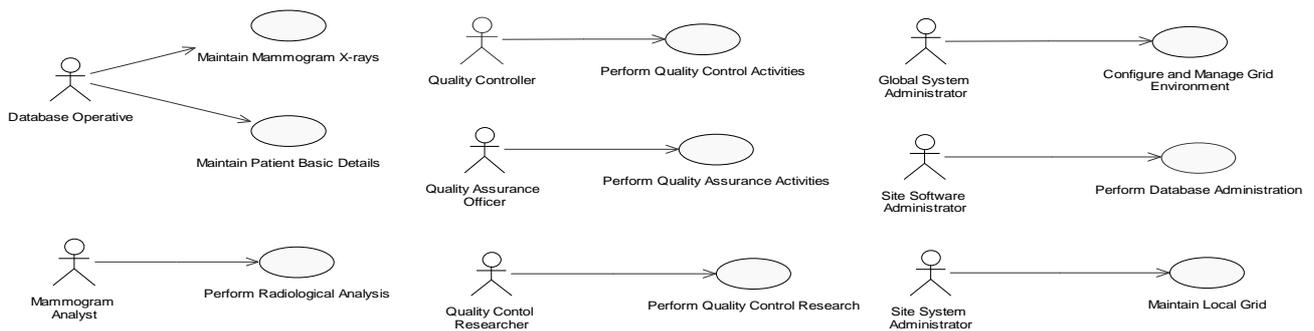

**Figure 2: MammoGrid's Main Use-Case View**

Furthermore, this use-case describes the tasks that the Quality Control Researcher may need to undertake such as installing local and Grid Algorithms for Acquisition Quality Control (AQC), and CAD.

*Use-Cases 7, 8 and 9:* System and Database Administration. The Global System Administrator actor configures and maintains the Grid infrastructure using the catalogue of sites within the MammoGrid environment to add to or modify the core software and global services. The Site System Administrator ensures that the MammoGrid system is accessed and operated only by designated users and that the definition of the access rights of these users is kept up to date. Finally, the Site Software Administrator maintains the MammoGrid database schema, updates and maintains the database authorization matrix.

## 5. Impact of the Use-Case Approach

The use-case model developed served as the common model (and vehicle) for communication between MammoGrid end-users, in particular physicians, and the development team. This approach proved its value in the communication and validation of user requirements in an incremental and controlled iterative manner.

Constraints on the requirements elicitation, specification and validation phases included the limited time available with domain experts, the geographic distance between the various stakeholders and hence the episodic nature of meetings. In the course of a visit of several days, domain experts could make themselves available for relatively frequent but rather brief meetings. Domain experts had no previous exposure to the kind of model used in software development. Software engineers on the project had little appreciation of the particular problems of mammography and breast cancer screening prior to this exercise. Moreover, the requirements team had to span the space between radiologists, Grid experts and medical image processing specialists, whether those working on the specification of the local workstation and those working on CADe.

The picture thus emerges of a triangular set of interactions in the development of these use-cases and associated documentation alongside them. Software engineers (from CERN and UWE) worked both with radiologists from Udine, Torino and Cambridge and with medical imaging engineers from Mirada and CADe specialists from Pisa and Sassari. A requirements engineering process has been applied to elicit and to analyze requirements, to develop requirements definitions and specifications, and to validate these. The process adopted reflected the stages of RUP, especially the requirements workflow in the inception and elaboration phases.

The main elicitation methods used were semi-structured interviews and strictly non-participant observation of medical procedures; with appropriate permissions, the team observed such procedures as basic mammography, ultrasound-guided biopsy, breast MRI, reading of a mammograms and other images, and X-ray examination of biopsy specimens. Based on these interviews and observations, the requirements team established an initial use-case model which was then iteratively and incrementally presented to radiologists at Udine (including those from Torino) and CADe experts from Pisa and Sassari, then to radiologists at Cambridge, then to imaging specialists and finally to two plenary meetings. In parallel, the team from Mirada constructed the 'acquisition system' data structures and thus helped validate the data requirements emerging from the use cases.

Since the two principal medical collaborators had slightly different aims in their involvement in the project, the incremental approach allowed each to absorb the requirements expressed by the other and to develop these into an increasingly sophisticated model of their professional practice. As a consequence, problem domain objects and relationships between them emerged and evolved, leading to a natural conceptual object model for MammoGrid. Consider the brief extract in figure 3 from the perform radiological use-case:

*Actor:*
    Mammogram Analyst
*Pre-Conditions*
    User-Authentication
*Non-functional requirements:*
    CADe Software Interface Requirements
*Flow of Events:*
    as per selection of the Mammogram Analyst to link to the appropriate extension point below
*Extension Points:*
    (1) View Mammogram and Patient Details
    (2) Annotate Mammograms and Patient Details
        (2.1) Diagnose Study
        (2.2) Diagnose Series
        (2.3) Annotate Image
        (2.4) Request CADe
        (2.5) Link Annotations
        (2.6) Request CADe in Mammogram Region
    (3) Execute Radiological Queries
        (3.1) Formulate Radiological Query
        (3.2) Refine Radiological Query
*Alternative Flows:*
    (1) Unsuccessful User Authentication
    (2) CADe Interface Error
    (3) Invalid Query Selections
*Post-Conditions*
    (1) Mammogram Image Annotated
    (2) Patient Details Changed
    (3) Results of Query Execution (Grid)
*Use-Case View:*

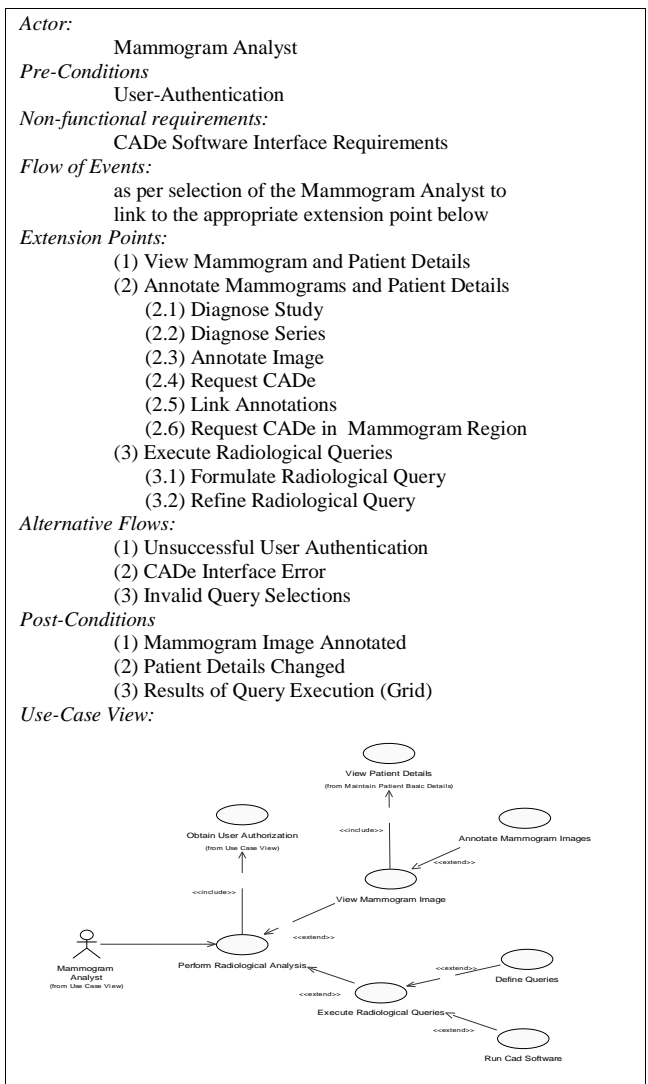

**Figure 3: Extract from Perform Radiological Analysis Use-Case.**

Data required to invoke a use-case, or to execute its actions, or to identify its subject, or to determine any further conditional actions, or to route its results, can be defined and structured into an appropriate class model. In the case of the MammoGrid project, this process was carried out in parallel with design of the Mirada Acquisition System (MAS), a highly specialized workstation and scanner combination with software to capture an appropriately digitized X ray image and the corresponding patient notes. In due course it must also support graphical annotation of the mammograms. The MAS provided a means to validate data requirements emerging from the use cases.

An analysis was also conducted of system non-functional requirements with identification and specification of those non-functional requirements that relate to a specific use-case as part of the overall specification of the use-case. While there has been heavy concentration on the use-case model and use-case specific non-functional requirements, it is worth mentioning that overall non-functional requirements did have its share of attention and analysis to specify those that are product-related, those which arise from organization and process, external constraints such as health and confidentiality, and interface specifications such as CADe and GridBox technical specifications. In this analysis constraints on the process of mammogram study, such as (but not limited to) usability, and security have been investigated and specified with some proper quantification to assess the adherence to these requirements at implementation time in addition to their impact on the selection and specification of the anticipated architecture of the MammoGrid.

Furthermore, we observe that a key feature of RUP, especially in the first phases (i.e. inception and elaboration), is that it is architecture-centric, which implies the addressing of software architecture as early as the inception phase with the aim to reach a stable software architecture towards the end of the elaboration phase. While this is not the subject of this paper in particular, one cannot leave the discussion here without stressing on the impact of addressing of software architecture and the early inclusion of software architects. The main issue here is the integrity of the use-case model presented in section 4. And, given that this project is Grid-centric, it was critical to have involvement of software architects that have been dealing with the Grid architecture to validate for example the use-case that deals with configuration and maintenance of the Grid environment.

## 6. Conclusions

This paper has shown a return on investment in use case modeling. This has manifested itself most importantly in the efficient communication with experts in two different but related domains: breast radiology and medical imaging. This experience suggests strongly that progress through the phases of the development can be enhanced through an iterative and incremental process of user requirements specification. This leads us to conjecture that the use-case modeling approach may result in reducing the cycle of the requirements engineering process in medium to large scale projects. This may be attributed to the role of the use-case approach in directing problem domain understanding, focusing elicitation, focusing analysis, better planning of iterations and their control, etc)

A logical architectural model for MammoGrid has also been derived from this requirements specification, identifying its main subsystems/components, the relationship between these and the specifications of their interfaces. Naturally, this paves the way for the more detailed design of the MammoGrid system. One important part of future work will be a systematic analysis of the use-case scenarios in the project with particular emphasis on how data flows between centres (e.g. hospitals) and to what extent a centralized or decentralized data management philosophy should be followed and on the choice between replicated data or replicated queries in the system.

There are relatively few published examples of the application of requirements engineering to medical domains and fewer of the application of UML and/or use-case modeling to medical applications. Among these we have noted the work of Bui et al, who have used a use case modeling approach to consider the requirements of evidence-based radiology. This work deserves closer attention and considered extension to a technology-rich environment in a separate work.

A further example is that of Teleservices and Remote Medical Care (TRMCS) [10] and Images and Diagnosis from Examples in Medicine (IDEM) [11]. In [10] the author takes a pragmatic if qualitatively critical approach and details various problems and deficiencies of UML, particularly concerning use case models and system decomposition. He proposes an alternative object-oriented design approach called ADORA (analysis and Design of requirements & Architecture) [12], which includes a hierarchy of abstract objects where each object 'truly integrates the aspects of structure, functionality, behavior and user interaction'. Experienced UML designers may take exception to some of the deficiencies identified in [10]. In particular, while the objections are real, one may take the view that a declarative reading of UML models is more appropriate than a procedural one, and that different UML models complement each other so as to address, if not actually resolve, these stringent criticisms.

In [11] UML is used to model the steps of a case-base reasoning system for medical image retrieval. The authors found that UML was particularly useful to display objects implementing their use cases of daily practices and to extend their object model to the level of a programming language. Whilst noting that a deeper analysis of the expressiveness of UML was required it was deemed sufficient for the medical image domain and useful in improving the communication between developers and users. This concurs with the findings of the research presented in this paper.

## 7. Acknowledgements

The authors take this opportunity to acknowledge the support of their home institutes and numerous colleagues responsible for the MammoGrid user requirements. The contributions of Professor Massimo Bazzocchi, Dr Ruth Warren and Dr Chiara del Frate are particularly acknowledged. In addition Professor Richard McClatchey wishes to acknowledge the support of the Royal Society in the preparation of this paper.

## References

[1] The Information Societies Technology project: "MammoGrid - A European federated mammogram database implemented on a GRID infrastructure". EU Contract *IST-2001-37614*

[2] The Rational Unified Process Model. See http://www.rational.com

[3] Alex A. T. Bui, Ricky K. Taira, John David N. Dionisio, Denise R. Aberle, Suzie El-Saden and Hooshang Kangarloo, Evidence-based Radiology: Requirements for Electronic Access, Academic Radiology 2002; 9:662–669

[4] I. Foster, C. Kesselman & S. Tueke., The Anatomy of the Grid – Enabling Scalable Virtual Organisations. *International Journal of Supercomputer Applications*, *15*(3), 2001.

[5] SMF : Mirada Solutions' Standard Mammogram Form. See http://www.mirada-solutions.com/smf.htm

[6] I. Sommerville, *Software Engineering,* (Addison-Wesley 2000)

[7] G. Kotonya and I. Sommerville, *Requirements Engineering* (Wiley 2000).

[8] G. Booch, J. Rumbaugh, and I. Jacobson ,*The Unified Modeling Language User Guide*. (Wesley Longman, Reading, MA, 1999).

[9] P. Krutchen, *The Rational Unified Process: An Introduction*, (Addison-Wesley, 1999)

[10] M Glinz, Problems and Deficiencies of UML as a Requirements Specification Language. *Proc. 10$^{th}$ IEEE Int Workshop on Software Specification and Design (IWSSD'00)*, San Diego, California. November 2000

[11] C LeBozec, M-C Jaulent, E Zapletal & P Degoulet, Unified Modeling Language and Design of a Case-Based Retrieval Systems in Medical Imaging. *Proc. AMIA'98 Annual Symposium: Accessing and Interpreting Digital Images*, Lake Buena Vista, Florida. November 1998.

[12] M Glinz, S. Berner, S. Joos, Object-oriented modeling with ADORA. *Information Systems, 27*( 6), 2002,  425-444. See: http://www.ifi.unizh.ch/groups/req/projects/ADORA